*Observation of a uniaxial ferroelectric smectic A phase*

Xi Chen[1], Vikina Martinez[1], Pierre Nacke[2], Eva Korblova[3], Atsutaka Manabe[4], Melanie Klasen-Memmer[4], Guillaume Freychet[5], Mikhail Zhernenkov[5], Matthew A. Glaser[1], Leo Radzihovsky[1], Joseph E. Maclennan[1], David M. Walba[3], Matthias Bremer[4], Frank Giesselmann[2], Noel A. Clark[1]*

[1]*Department of Physics and Soft Materials Research Center,*
*University of Colorado, Boulder, CO 80309, USA*

[2]*Institute of Physical Chemistry, University of Stuttgart, 70569 Stuttgart, Germany*

[3]*Department of Chemistry and Soft Materials Research Center,*
*University of Colorado, Boulder, CO 80309, USA*

[4]*Electronics Division, Merck KGaA, 64293 Darmstadt, Germany*

[5]*Brookhaven National Laboratory, National Synchrotron Light Source-II*
*Upton, NY 11973, USA*

*Abstract*

We report the smectic $A_F$, a new liquid crystal phase of the ferroelectric nematic realm. The smectic $A_F$ is a phase of small polar, rod-shaped molecules which form two-dimensional fluid layers spaced by approximately the mean molecular length. The phase is uniaxial, with the molecular director, the local average long-axis orientation, normal to the layer planes, and ferroelectric, with a spontaneous electric polarization parallel to the director. Polarization measurements indicate almost complete polar ordering of the ~10 Debye longitudinal molecular dipoles, and hysteretic polarization reversal with a coercive field ~2 x $10^5$ V/m is observed. The SmA$_F$ phase appears upon cooling in two binary mixtures of partially fluorinated mesogens: 2N/DIO, exhibiting a nematic (N) – smectic $Z_A$ (SmZ$_A$) – ferroelectric nematic (N$_F$) – SmA$_F$ phase sequence; and 7N/DIO, exhibiting an N – SmZ$_A$ – SmA$_F$ phase sequence. The latter presents an opportunity to study a transition between two smectic phases having orthogonal systems of layers.

*Significance*

Liquid crystal science grows in richness and applicability with each new phase that is found or created. The recent discovery of the ferroelectric nematic was both thrilling and unexpected, since it appeared in new molecules not much different in structure from many similar materials studied over the last 100 years. Clearly, significant, and sometimes seemingly magical, secrets remain to be discovered in the complexities of organic molecular architecture and interaction. A fundamental question following the ferroelectric nematic discovery was whether there could also be a ferroelectric smectic A, the nematic-companion phase obtained when molecules spontaneously slide to form planar, fluid layers normal to their molecular long axes. Here we report such a phase, broadening the ferroelectric nematic realm.

## INTRODUCTION

Proper ferroelectricity in liquids was predicted in the 1910's by P. Debye [1] and M. Born [2], who applied the Langevin-Weiss model of ferromagnetism to propose a liquid-state phase change in which the ordering transition is a spontaneous polar orientation of molecular electric dipoles. A century later, in 2017, two groups independently reported, in addition to the typical nematic (N) phase, novel nematic phases in strongly dipolar mesogens, the "splay nematic" in the molecule RM734 [3,4,5] and a "ferroelectric-like nematic" phase in the molecule DIO [6]. These nematic phases were subsequently demonstrated to be ferroelectric in both RM734 [7] and in DIO [8,9], and to be the same phase in these two materials [9]. This new phase, the ferroelectric nematic ($N_F$), is a uniaxially symmetric, spatially homogeneous, nematic liquid having ≳90% polar ordering of its longitudinal molecular dipoles [7,9]. A related new phase recently observed is the helical ferroelectric $N_F$ [10,11,12,13,14], obtained by chiral doping of RM734, DIO, or their homologs, or by introducing chiral tails into the molecular structures [15]. DIO also exhibits an additional phase, found between the N and $N_F$ [6], which we have recently characterized, terming it the smectic $Z_A$ [16], and showing it also to be new: a density-modulated antiferroelectric exhibiting lamellar order with ~18 nm repeats, comprising pairs of ~9 nm-thick layers with alternating polarization, the director and polarization being oriented parallel to the layer planes.

Here we introduce another new phase of the ferroelectric nematic realm, the smectic $A_F$, a uniaxial, lamellar phase with the director normal to the layers and a spontaneous polarization along the director. Schematic drawings of the phases discussed here, sorted into macroscopically non-polar and polar types, are shown in *Fig.1*, along with the molecular structures and phase sequences of the mesogens used in the mixtures. The macroscopically non-polar, paraelectric nematic (N) and smectic A (SmA) phases, the ferroelectric nematic ($N_F$) and ferroelectric smectic A (Sm$A_F$) phases, and the antiferroelectric Sm$Z_A$ phase are sketched in *Fig. 1A*, the light-to-dark shading of the schematic molecules indicating their dipolar symmetry. The Sm$A_F$ phase is observed in 50:50 wt% AUUQU2N/DIO (2N/DIO) and AUUQU7N/DIO (7N/DIO) mixtures. The region of *Fig. 1A* shaded yellow shows the generic phase sequence observed in the mixtures on cooling (Iso → N → Sm$Z_A$ → $N_F$ → Sm$A_F$ → X), noting that some phases may be missing in a given component or mixture. For example, none of the single components exhibits the Sm$A_F$ phase, and the 7N/DIO mixture does not have the $N_F$ phase. The first mesophase that appears on cooling any of the components and mixtures from the isotropic is the conventional dielectric nematic (N) phase, which, in the present context, is also considered paraelectric. They all cool from the N into the antiferroelectric smectic Z (Sm$Z_A$) [16] phase.

The 2N/DIO mixture then transitions first to the $N_F$ phase and then, on further cooling, to the Sm$A_F$, while 7N/DIO goes directly to the Sm$A_F$. This enables a comparative study of both the $N_F$ → Sm$A_F$ and Sm$Z_A$ → Sm$A_F$ transitions, the latter featuring the simultaneous disappearance of the SmZ$_A$ layering parallel to the director and the formation of the SmA$_F$ layering normal to the director, in the absence of any director/polarization reorientation.

In contrast to the conventional dielectric smectic A phase, the ferroelectric smectic A phase exhibits a macroscopic polarization **P**, with the polarization in every layer pointing in the same direction, along the director, **n**, normal to the layer planes. The phase is uniaxial and has a high degree of polar order (polar order parameter $p > 0.9$). Domains of opposite polarization separated by polarization-reversal walls (sketched in **Fig. 1A**) are observed in regions with continuous smectic layering .

This ferroelectric phase is distinct from the phases previously described in several families of uniaxial "polar smectics", including the monolayer paraelectric SmA$_1$, the partial bilayer SmA$_d$, the antipolar bilayer SmA$_2$ phase, and a variety of polarization-modulated phases (Sm, Sm, etc.) of dipolar molecules [17,18,19], in that these all have zero net average polarization [20]. Tournilhac and co-workers claimed initially to have observed macroscopic polarization normal to the layers in a small-molecule, smectic A phase, based on evidence of piezoelectricity and non-linear dielectric behavior [21,22] but their subsequent x-ray scattering study revealed a smectic unit cell-doubling [23], leading to the conclusion that the phase in question was a bilayer smectic of the SmA$_d$ variety, and that the observed electrical effects were manifestations of bilayer antiferroelectricity. The SmA$_F$ is also different from the orthogonal polar smectic phases exhibited by some bent-core mesogens, which form biaxial smectics with the spontaneous polarization oriented parallel to the smectic layers [24, 25,26].

### _RESULTS_

_X-ray scattering_ – We have previously carried out X-ray diffraction, polarized light microscopy, and polarization measurement studies of the single molecular components, DIO [9,16] and 2N,7N [27] shown in **Fig. 1B**. Here we focus on the binary mixtures 2N/DIO and 7N/DIO. All of our observations indicate that the N, N$_F$ SmZ$_A$, and SmA$_F$ phases observed in these different single components and/or in the mixtures exhibit common experimental characteristics and appear, respectively, to be the same phases in the different materials: the N phases are homogeneous, uniaxial nematics, the N$_F$ phases are homogeneous, uniaxial nematics with a macroscopic polarization along the nematic director, and the SmZ$_A$ is the same bilayer antiferroelectric phase in all of the components and mixtures, with a layer spacing $d_M \approx 90$Å in DIO, $d_M \approx 81$Å in the 2N/DIO mixture, and $d_M \approx 60$Å in the 7N/DIO mixture. The period of the layer-by layer antiferroelectric polarization alternation is $2d_M$.

In this study we describe the LC behavior of 50:50% 2N/DIO and 7N/DIO mixtures, both of which exhibit the SmA$_F$. We find that these mixtures show: (_i_) similar SAXS from the SmA$_F$ layering, with smectic layer spacing close to the mean molecular length; (*ii*) similar uniaxial bi-refringence; (*iii*) similar SmA-like optical textures; (*iv*) similar response of the SmA$_F$ to surface alignment conditions and applied electric field; and (*v*) similar SmZ$_A$ and SmA$_F$ polarization reversal dynamics. We discuss the two mixtures separately because of the differences in how the SmA$_F$ grows in on cooling, 2N/DIO coming from the N$_F$ phase and 7N/DIO coming from the SmZ$_A$ phase, as this condition strongly affects the textural morphology of the SmA$_F$.

For the SAXS and WAXS experiments, the mixtures were filled into 1 mm diameter, thin-wall capillaries and the director ***n*** (yellow arrow in ***Fig. 2A***) was aligned by an external magnetic field ***B*** (red arrow). The SAXS and WAXS was nonresonant, with diffraction images of the samples obtained in transmission on the SMI beamline (12-ID) at NSLS II, a microbeam with an energy of 16.1 keV and a beam size of 2 μm x 25 μm.

<u>2N/DIO</u> – Typical SAXS and WAXS images obtained on cooling the 50:50% 2N/DIO mixture from the N$_F$ to the SmA$_F$ phase are shown in ***Fig. 2A***. In the N$_F$ phase at $T$ = 57.9°C, we observe a nematic-like, diffuse scattering arc peaked in azimuthal orientation with scattering vector ***q*** along ***n***, coming from the head-to-tail pair correlation of the molecules along ***n***. Line scans of the scattering intensity through these peaks are shown in ***Fig. 2B***. As seen in the inset of ***Fig. 2B***, the SmA$_F$ phase is heralded by the appearance of a new, resolution-limited peak along $q_z$, first showing up at $T \approx 56$ °C, at $q_{zAF} \approx 0.267$ Å$^{-1}$, a wavevector very close to the diffuse nematic peak at $q_z \approx 0.271$ Å$^{-1}$. This behavior indicates a first-order phase transition from the N$_F$ to the SmA$_F$, in accord with our polarized light microscope observations. The corresponding layer spacing is $d_{AF}$ = 23.5Å, comparable to the concentration-weighted average molecular length of DIO (23.2 Å) and 2N (23.4 Å). The absence in the SAXS images of half-order peaks at $q_z = q_{zAF}/2$ indicates that there is no observed tendency for bilayer fluctuations or ordering in the SmA$_F$ in this mixture. The WAXS diffraction image in ***Fig. 2A*** shows the second-harmonic scattering from the layers at $2q_{zAF} \approx 0.53$ Å$^{-1}$. The full width at half-maximum azimuthal mosaic distribution of ***n*** in the mag-netically aligned sample is initially ~5°. The scattering pattern rotates in the SmA$_F$ phase on cooling due to dynamical textural rearrangements in the capillary [16] and at lower temperature there is some detectable scattering from the layering at all azimuthal angles as the magnetic torque is not strong enough to maintain the alignment of the increasingly rigid smectic layers.

<u>7N/DIO</u> – Typical SAXS diffraction images obtained on cooling the 50:50% 7N/DIO mixture from the SmZ$_A$ to the SmA$_F$ phase are shown in ***Figs. 3A*** and ***4***. The SmA$_F$ scattering is qualitatively similar to that of the 2N/DIO mixture. In the SmZ$_A$ phase at $T$ = 43.6°C, the SAXS shows a diffuse, nematic-like scattering arc, peaked with scattering vector ***q*** along ***n***, coming from head-to-tail pair correlations of the molecules along ***n***||***z***. Radial line scans of the scattering intensity along ***n*** (the white lines depicted in ***Fig. 3A***) are shown in ***Fig. 3B***.

As in the 2N/DIO mixture, the SmA$_F$ phase is characterized by a new, resolution-limited peak along $q_z$, first appearing at $T \approx 31$ °C, at $q_{zAF} \approx 0.245$ Å$^{-1}$, at the maximum of the diffuse nematic peak, as shown in the inset of *Fig. 3B*. The corresponding layer spacing $d_{AF} = 25.6$ Å is comparable to the concentration-weighted average molecular length of DIO (23.2 Å) and 7N (29.1 Å). The absence in the SAXS images of half-order peaks at $q_z = q_{zAF}/2$ again indicates that there is no tendency to form bilayers. As in the DIO/2N mixture, the scattering pattern rotates in the SmA$_F$ phase due to dynamical textural rearrangements in the capillary with changing temperature [16]. The scattering arc becomes wider in the SmA$_F$ as the effectiveness of the magnetic field alignment is reduced on cooling.

Finally, the equatorial Bragg spots at $q_y = q_{yM}$ coming from the density modulation due to the smectic layering of the SmZ$_A$, which are observed in both the 2N/DIO and 7N/DIO mixtures but are not visible in *Figs. 2A* or *3A* because they are relatively weak, are shown in *Fig. 4*.

*__Polarized optical transmission microscopy__* enables direct visualization of the director field, $\boldsymbol{n}(\boldsymbol{r})$, and, apart from its sign, of $\boldsymbol{P}(\boldsymbol{r})$. These observations provide key evidence for the macroscopic ferroelectric ordering, uniaxial optical textures, and fluid layer structure of the SmA$_F$ phase of the 2N/DIO and 7N/DIO mixtures.

*7N/DIO* – The 50:50% 7N/DIO mixture was studied in an $d = 3.5$ μm cell with anti-parallel surface rubbing (an antipolar cell) with planar electrodes on one surface separated by a 1 mm gap. In the N phase, the LC formed a uniformly aligned monodomain with $\boldsymbol{n}$ along the buffing direction, as previously observed in the N phase of DIO [16]. In the 7N/DIO mixture with no field applied there is little change in sample appearance with temperature of these cells, the nematic texture being maintained upon cooling into the SmZ$_A$ and SmA$_F$ phases, as seen in *Figs. 3C*1,2. At the SmZ$_A$ to SmA$_F$ transition, the SmZ$_A$ layers parallel to $\boldsymbol{n}$ disappear while new SmA$_F$ layers, normal to $\boldsymbol{n}$, form. The birefringence color is a uniform blue-green everywhere in the cell and changes only slightly during the N $\rightarrow$ SmZ$_A$ $\rightarrow$ SmA$_F$ cooling sequence, providing evidence that the phase is uniaxial or only weakly biaxial and that the optical anisotropy is nearly the same in all three phases. The uniaxiality of the N phase and the weak biaxiality of the SmZ$_A$ have been demonstrated previously [16].

The SmZ$_A$ layers adopt bookshelf geometry, with the smectic layers normal to the plates and with Rapini–Papoular type anchoring of the molecules aligning the director along the rubbing direction. The transition of the antiferroelectric SmZ$_A$, with its layer-by-layer alternation of $\boldsymbol{P}$, to the ferroelectric SmA$_F$ phase is achieved by a coarsening process in which layers with the same sign of $\boldsymbol{P}$ coalesce into broader stripes of uniform polarization extended along $z$, leading to a texture of irregular, needle-like ferroelectric domains of alternating polarization in the SmA$_F$.

While this process produces only subtle changes in the textures in the absence of applied field (compare **Figs. 3C**1 and 2), application of an in-plane electric field normal to **n** induces rotation of **P** in opposite directions in domains with opposite polarization, facilitating and inducing the coarsening of the domain pattern (**Figs. 3C**3 to 6). This electric field response becomes increasingly dramatic as the stripes coarsen from the nanoscale to the microscale.

After extended application of weak electric fields, the SmA$_F$ cell anneals, in the absence of further applied field, into long, rectangular bookshelf domains with uniform birefringence and excellent extinction, typical of weakly oriented smectic A textures, as shown in **Figs. 3D**1,2. Sufficiently large transverse DC fields can completely reorient the SmA$_F$ layers so that **P** and **n** become aligned along **E**, normal to the buffing direction (**Figs. 3D**5). In the N$_F$ phase, this kind of global, field-induced reorientation is essentially thresholdless, reversing readily on applied field reversal, but in the SmA$_F$ phase there is a distinct threshold for switching and hysteresis in the response, manifest in the polarization data of **Fig. 6**. This behavior can be understood by considering that field-induced reorientation of a spatially uniform SmA$_F$ can only by accommodated by the generation of a population of gliding edge-dislocations, an inherently non-linear process. The effect of this threshold is immediately apparent in the electro-optic behavior in cells with in-plane electrodes. In an applied field, these domains reorient, buckle, and, in sufficiently large applied field, coarsen to form large domains with the **n, z,** and **P** all oriented along the field, normal to the buffing direction (**D**3 to 5). Thus it appears that, during field-induced reorientation, **n, z,** and **P** remain coupled together, with the threshold originating from the elasticity and plasticity of the smectic layering. This threshold also results in the appearance of a coercive field in the polarization hysteresis (**Fig. 6**). The N$_F$ typically responds readily to in-plane applied electric fields present anywhere in the cell, including above metal or ITO electrodes, and even to small fringing fields far from any electrodes. In the SmA$_F$ phase, in contrast, this response becomes sub-threshold and is eliminated from these peripheral areas, with electro-optic effects confined to the designated active areas of the cell where the field is strongest, as seen in **Figs. 3D3** to **3D5**.

An interesting side observation is the lack of field response in the regions to the left and right sides of the air bubbles in **Figs. 3C**3 to 6. This "shadowing" effect is a direct consequence of the ferroelectric nature of SmA$_F$ phase. The air bubble in the middle of the gap between the electrodes into a series connection of impedances: the left/right electrode and right/left boundary of the bubble with SmA$_F$ as filling medium and the air bubble with air as the filling medium. The regions with SmA$_F$ as a medium have low electrical impedance due to reorientation the large polarization density [28,29], while the air bubble capacitance will be small, dropping most of the applied voltage, and leaving little field response in the adjacent LC.

_2N/DIO_ – The 50:50% 2N/DIO mixture was studied in an antipolar  $d$ = 3.5 μm cell (with anti-parallel surface rubbing), and in a synpolar $d$ = 5 μm cell (with parallel surface buffing).

In the antipolar cell, the surface anchoring imposes a twist structure in the N$_F$ phase in which the director/polarization field $n(r)$,$P(r)$ rotates by π through the thickness of the cell [11,10]. The twisted N$_F$ state appears from pinkish to blueish in **Fig. 2C**, which shows the cell being cooled through the first-order N$_F$ to SmA$_F$ transition, with blue-green SmA$_F$ domains growing in the upper part of the field of view. The uniformity of the birefringence color, and the observation that the SmA$_F$ domains can be rotated to extinction between crossed polarizers, indicate that the director twist has been expelled and the principal optic axis along $n$ is locally uniform through the cell in the SmA$_F$ regions, with $n(r)$ uniformly parallel to the plates. The growing SmA$_F$ domains are not strongly orientationally aligned by the cell surfaces initially, most likely because of the ambivalence of these now polar domains towards the antipolar surfaces. The effects of applying a weak, probe electric field normal to the director are shown in **Figs. 5A-C**, confirming that each domain is internally homogeneously polar (black/white arrows) with orientation along the local director, some pointing up and some pointing down. The expulsion of bend and twist of $n(r)$ by the smectic A$_F$ layering, and expulsion of splay of $n(r)$ in order to eliminate polarization charge, results in steady-state textures of uniformly oriented SmA$_F$ blocks, as shown in **Fig. 2D**, in which there are distinct domain boundaries running either parallel or perpendicular to $n$. The boundaries parallel to $n$ (approximately vertical in these images) are polarization-reversal walls like those found in the N$_F$ phase [7], while those perpendicular to $n$ are either melted grain boundaries of the type commonly found in SmA phases not completely aligned by weak buffing [30], or are polarization-stabilized kinks (PSKs) [31], as sketched in the inset of **Fig. 2E**. Changes in the sign of $P(r)$ across the horizontal boundaries would generate maximal space-charge and are thus avoided, with jumps in the orientation of $P(r)$ at these locations being limited to 10° or less. In general, there is a tendency to form long SmA$_F$ domains of uniform polarization extended along the director, as seen in **Figs. 2D** and **3D**. The internal variation of orientation within the blocks is generally only a few degrees and tends to be manifested as bend of the director field, which must be mediated by edge dislocations in the SmA$_F$ layering.

More detailed structures of the transition regime that mediates the growth of the uniform SmA$_F$ domains into the twisted region are shown in **Figs. 5D-F**. Here remnant diamond-shaped N$_F$ twist domains connect to surrounding uniform SmA$_F$ domains by forming PSK domain boundaries with the polarization directions in the sample midplane indicated in **Fig. 5E**. More detailed images of the transition regime that mediates the growth of the uniform SmA$_F$ domains into the twisted N$_F$ region are shown in **Figs. 5D-F**. Here remnant diamond-shaped N$_F$ twist domains connect to surrounding uniform SmA$_F$ domains by forming PSK domain boundaries with the polarization directions in the sample mid-plane indicated in **Fig. 5E**. Similar structures constitute the zig-zag SmA$_F$ – N$_F$ boundary line seen in **Fig. 5F**.

If the SmA$_F$ is heated back into the N$_F$ phase, the removal of the layering constraints enables the polarization-reversal walls to restructure into nematic splay-bend walls [7] extended along the director, separated by areas of uniform polarization, as seen in **Figs. 2D**2,3). The horizontal melted grain boundaries disappear in the absence of layering, while the horizontal PSK lines can persist into the N$_F$ but then also melt away, leaving only the splay-bend walls (bright lines in **Figs. 2D**2,3). Because of the antiparallel boundary conditions, the initially uniform N$_F$ states are only metastable and the inherently twisted cores of the splay-bend walls act as nucleation sites for the formation of lower-energy, twisted domains, which eventually spread to cover the entire area (**Figs. 2D**5).

In the synpolar cell, the surface treatment stabilizes monodomains in which $n$ is homogeneously aligned along the buffing direction. The texture and birefringence of these monodomains barely change on cooling through the N – SmZ$_A$ – N$_F$ – SmA$_F$ phases, exhibiting excellent extinction between crossed polarizers in the N$_F$ and SmA$_F$ phases except near air bubbles, as seen in **Fig 2E**. The first image shows how the uniform background N$_F$ director field favored by the cell surfaces is distorted to accommodate the non-uniform $n(r)$ orientation imposed by the boundary conditions at the bubble boundaries, where the $n(r)$ field is tangential, a configuration which requires only bend of the director and minimizes the amount of space-charge deposited at the LC/air interface. On the sides of the bubble, the director field distortion relaxes continuously with distance, with the director field eventually becoming indistinguishable from the surrounding uniform state. At the top and bottom of the bubble, however, the 90° angular mismatch of the circumferential $P(r)$ and the uniform background is accommodated by a "fracture" of $P(r)$ in the form of a polarization-stabilized kink [31], sketched in the inset. The PSK has a minimum-energy discontinuity in $P(r)$, with an internal structure determined by the balance of Frank elastic and electrostatic interactions, the latter manifested as an attraction between sheets of polarization charge of opposite sign (red and green in the inset), which stabilizes the wall. The kink orientation locally bisects the angle between the incoming and outgoing $P(r)$ directions, leading to a globally parabolic boundary between the regions with uniform and circular bent-director fields having minimal bulk polarization charge. Such 2D parabolic textures are readily observed in N$_F$ cells in which $P(r)$ is parallel to the bounding plates, its typically preferred orientation.

At the N$_F$ – SmA$_F$ transition, the areas of uniform director orientation expand, a result of the appearance of the SmA layering. In the absence of edge and screw dislocations, smectics expel both bend and twist of $n(r)$, allowing, in inhomogeneously aligned non-polar smectics A, layering defects only in the form of focal conic domains, as these require only splay of $n(r)$. However, in the polar SmA$_F$ phase, splay is also suppressed because of the associated polarization charge, leading to a strong tendency to form domains of uniform $\boldsymbol{n(r)}$.  As the smectic layers form on cooling, the bent-director region near the bubble, in which there is both bend and twist of $\boldsymbol{n(r)}$, is therefore reduced in size,  as shown in the second image of **Fig 2E**.  The remaining bent- and twisted-director region near the bubble must be accommodated by edge and screw dislocations in the smectic A layering.

**_Polarization dynamics and field-induced phase transitions_** – The polarization was measured in a $d = 17$ μm ITO-sandwich cell with bookshelf layering using an low-frequency (8 Hz), 30 V peak amplitude triangle wave. The electrical response of the 2N/DIO mixture is summarized in **Fig. 6**. At the beginning of the current voltage cycle shown in **Fig. 6A**, the applied voltage is large and negative ($V(t) \approx$ -30 V), at which time any ions have been pulled to the cell surfaces.  In the N phase ($T > 84^{\circ}$C), the current shows a bump following the sign change of $V(t)$, which we attribute to ions  This current is subtracted out when calculating $\boldsymbol{P}$.  In the SmZ$_A$ phase ($84^{\circ}$C $> T > 68^{\circ}$C), LC repolarization peaks appear when the voltage is decreasing, growing in area, with their peak center voltages $V_{FA}$ becoming smaller on cooling, behavior very similar to that of neat DIO (Fig. 5 in [7]).  This is typical antiferroelectric behavior, the peaks marking the return at finite voltage of the field-induced ferroelectric state to the antiferroelectric ground state.  In the SmZ$_A$ the polarization current interacts with ion current in a complex way following each sign change of $V(t)$, so $P(T)$ is obtained by doubling the $I(t)$ area left of the pink line, where there is no ion current. In the N$_F$ phase the Goldstone-mode mediated reorientation and reversal of $\boldsymbol{P}$ produces the current peak at the zero crossing of $V(t)$, followed by an ion peak for $t > 0$.  $P(T)$, taken as the area of the big peak, is found to be comparable to that of neat DIO.  In the SmA$_F$ phase, the ion current disappears altogether and polarization reversal occurs after the zero-crossing, at a finite voltage corresponding to the coercive field $E_c$ plotted as solid symbols in **Fig. 6B** and shown schematically in the adjoining hysteresis loop.

**_DISCUSSION_**

The discovery of the ferroelectric smectic A phase adds an exciting new dimension to the ferroelectric nematic realm The ferroelectric nematic, chiral ferroelectric nematic, and antiferroelectric smectic Z$_A$ have each opened unanticipated doors to new soft matter science and technology, and here the smectic A$_F$ joins in this development.  The SmA$_F$ is a layered, spontaneously polar fluid, the long-sought-after proper ferroelectric smectic A liquid, its reorientable macroscopic spontaneous polarization now definitively proven.  The transitions to the SmA$_F$, either N$_F$ to SmA$_F$ or SmZ$_A$ to SmA$_F$ are first-order, and rather subtle in cells with parallel polar surface anchoring with their textures and many of their phase properties exhibiting continuity through the transition.  The polarization, ~90% saturated in the N$_F$, remains so in the SmA$_F$, in the presence of the long-range side-by-side molecular positioning implied by the smectic A layer ordering.

This is something of a conundrum since side-by-side is the highest energy arrangement of similarly oriented dipoles.  Atomistic simulation promises to lead to a clearer understanding of how such a combination can exist in a liquid.

<u>*MATERIALS AND METHODS*</u>

The mixtures were studied using standard liquid crystal phase analysis techniques described previously [7,10,16], including polarized transmission optical microscopy observation of LC textures and their response to electric field, x-ray scattering (SAXS and WAXS), and techniques for measuring polarization and determining electro-optic response.

<u>*Materials*</u> – DIO**,** shown in ***Fig. 1*** and first reported in Ref. [6], was synthesized for these experiments as described in [9].  Synthesis of AUUQU2N and AUUQU7N in ***Fig. 1*** followed that of AUUQU3N, described in [32].

<u>*X-ray scattering*</u> – For SAXS and WAXS, the LC samples were filled into 1 mm-diameter, thin-wall capillaries. The director ***n*** was aligned with an external magnetic field normal to the beam.  Diffraction data presented here were obtained on the SMI beamline at NSLSII with a photon energy of 16 keV (wavelength = 0.775 Å).  At this wavelength, the desired range of scattering vectors ($q <$ 0.5 Å$^{-1}$) encompasses a small range of scattering angles ($\theta < 3°$), so that the Ewald sphere can be approximated as an Ewald plane, ($q_y$,$q_z$), which is normal to the beam, with ***z*** along the magnetic field ***B*** and director ***n*** orientation.  SAXS and WAXS images of 2N, 7N, and their mixtures with DIO obtained on cooling from the Iso to the nematic phase, show the intense, diffuse scattering features at $q_z \sim 0.25$ Å$^{-1}$ and $q_y \sim 1.4$ Å$^{-1}$ from end-to-end and side-by-side molecular positional pair correlations, respectively, that are characteristic of this type of polar mesogen.

<u>*Electro-optics*</u> – For making electro-optical measurements, the mixtures were filled into planar-aligned, in-plane switching test cells with unidirectionally buffed alignment layers on both plates. Cells with antiparallel buffing on plates separated by $d = 3.5$ μm, and with parallel buffing on plates with a $d = 5$ μm separation, were used. In-plane ITO electrodes were spaced by 1 mm and the buffing was parallel to this gap.  Such surfaces give a quadrupolar alignment of the N and SmZ$_A$ directors along the buffing axis and polar alignment of the N$_F$ at each plate.  Antiparallel buffing stabilizes a twisted configuration in the N$_F$ phase, generating a director/polarization field that is parallel to the plates and undergoes a π-twist between the plates [10].  Parallel buffing generates polar monodomains in the N$_F$ and SmA$_F$ phases.

<u>*Polarization measurement*</u> – We measured the *I(t)–V(t)* characteristics of the 50:50 wt% 2N/DIO mixture as a function of temperature for AC electric field applied along ***n.***  The current response *I(t)* to an 8 Hz, 30 V peak amplitude triangle wave *V(t)* was measured in a $d = 17$ μm ITO-sandwich cell with bookshelf layering during an N → SmZ$_A$ → N$_F$ → SmA$_F$ cooling scan.

## *ACKNOWLEDGEMENTS*

This work was supported by NSF Condensed Matter Physics Grants DMR-1710711 and DMR-2005170, by Materials Research Science and Engineering Center (MRSEC) Grant DMR-1420736, by University of Colorado Lab Venture Challenge OEDIT Grant APP-354288, and by Polaris Electro-Optics. This research used the microfocus Soft Matter Interfaces beamline 12-ID of the National Synchrotron Light Source II, a U.S. Department of Energy (DOE) Office of Science User Facility operated for the DOE Office of Science by Brookhaven National Laboratory under Contract No. DE-SC0012704. X-ray experiments were also performed in the Materials Research X-Ray Diffraction Facility at the University of Colorado Boulder (RRID: SCR_019304), with instrumentation supported by NSF MRSEC Grant DMR-1420736.

*FIGURES*

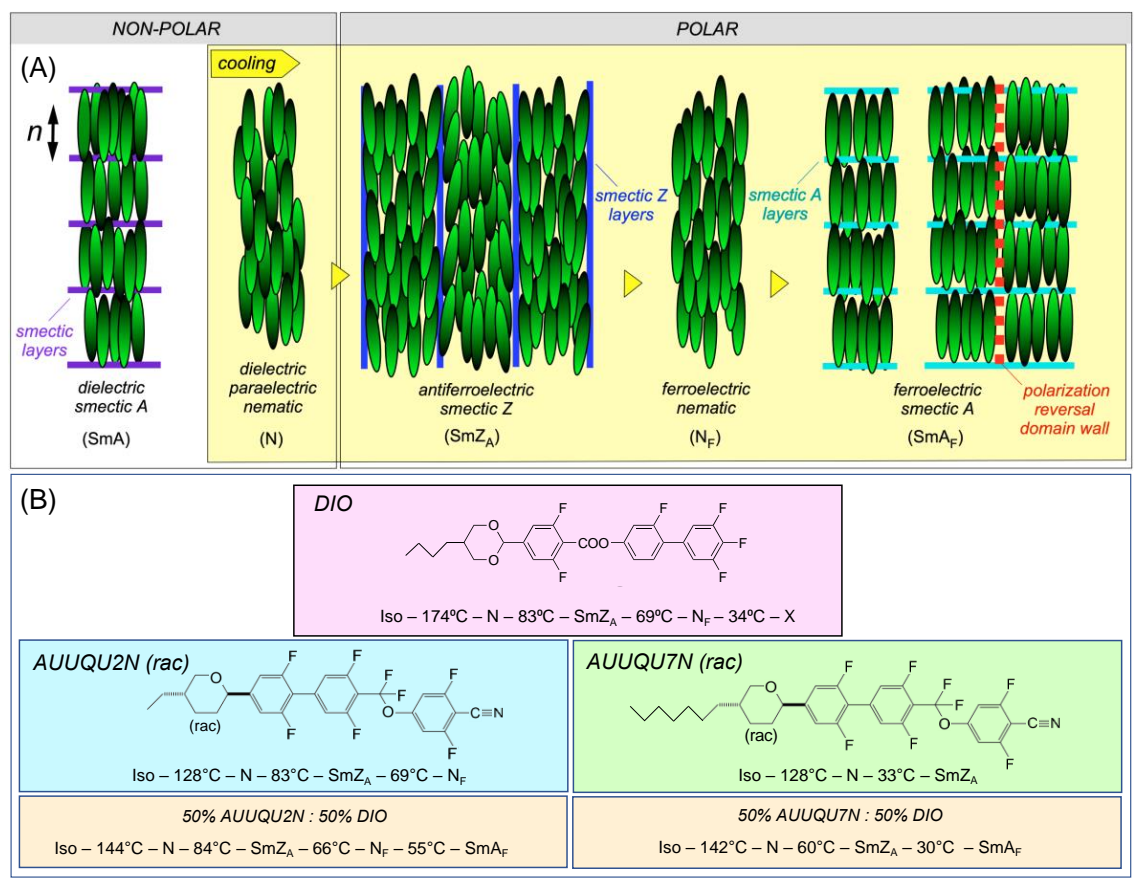

*Figure 1*: Structures, phase sequences and schematics of the liquid crystal phase behavior of 2N, 7N, and DIO single components, and their indicated mixtures. The relevant phases of rod-shaped molecules with on-axis electrical dipole moments are shown, where the dipole direction of a schematic molecule is indicated by its black-to-green shading. (*A*) Sketches of the phase organization, grouped into macroscopically non-polar and polar types. The experiments reported here confirm the existence of the previously described paraelectric nematic (N) [3,4], antiferroelectric smectic Z (SmZ$_A$) [16], and ferroelectric nematic (N$_F$) [4,5,6,7] phases, as well as the new SmA$_F$ phase. These phases appear upon cooling with the general order vs. *T* shown in the yellow-shaded area. Note that the N$_F$ phase is missing in the 7N/DIO mixture, allowing for a direct smectic Z to smectic A$_F$ transition. The solid, heavy lines depict smectic layering. The SmA$_F$ phase is spontaneously ferroelectric, with polarization P ~ 6 μC/cm$^2$ and polar order parameter *p* ~ 0.9, values comparable to those of the N$_F$ phase of DIO [6] and RM734 [7]. Polarization reversal is effected by the motion of pure polarization reversal domain walls. The antiferroelectric layer-by-layer alternation of polarization induces director splay modulation in the SmZ$_A$ phase, but splay is suppressed in the ferroelectric N$_F$ and SmA$_F$ phases.

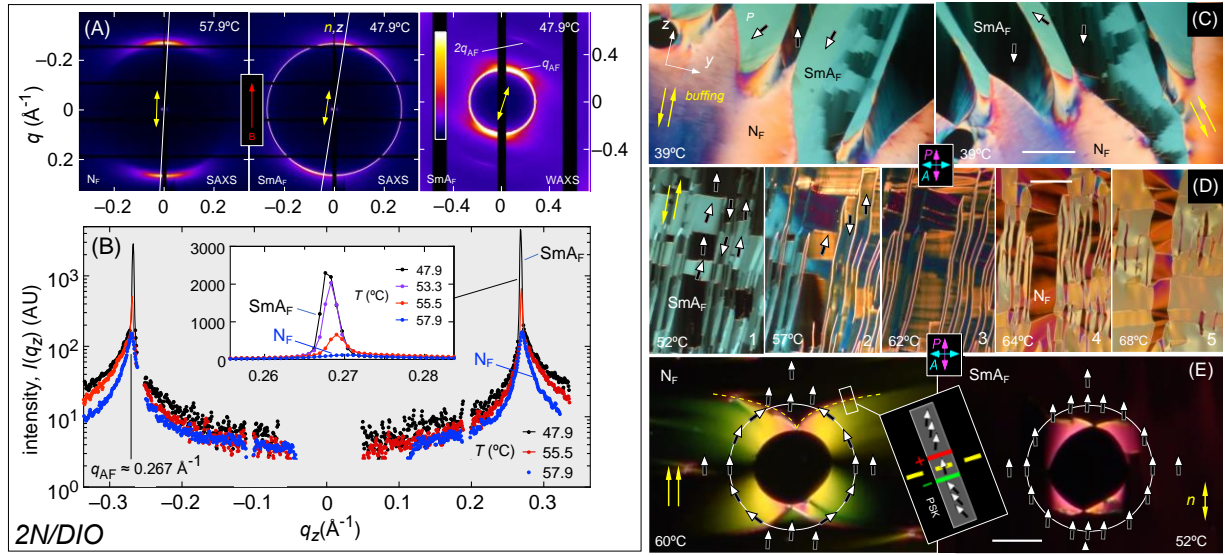

***Figure*** **2**: X-ray scattering and polarized microscopy textures of the N_F and SmA_F phases in the 50:50% 2N/DIO mixture. (**A**) Typical non-resonant SAXS and WAXS obtained on cooling from N_F to SmA_F. In the N_F phase at 57.9 °C, there are nematic-like diffuse scattering arcs, peaked along **n** at $q_z \sim 0.27$ Å$^{-1}$, coming from head-to-tail correlation of the mixture molecules. (**B**) Radial intensity scans along the **n**,$q_z$ direction (the white lines in (A)) at different temperatures. In this experiment, the scattering pattern rotates azimuthally by as much as 10° due to textural reorganization within the capillary as the smectic layers form. The initially diffuse smectic peak sharpens somewhat on cooling, until a distinct, resolution-limited SmA_F Bragg reflection appears in the **n** direction at $T \approx 56$°C, as shown in the inset, indicative of smectic ordering with the layer planes normal to **n**. The scattering vector $q_{zAF} \approx 0.267$ Å$^{-1}$ corresponds to a SmA_F layer spacing of 23.5 Å, close to the wt% average molecular length of DIO (23.2Å) and 2N (23.4Å). The SmA_F peak position is very close to that of the nematic peak, as expected for an orthogonal smectic phase. The polarized light microscope images show the 50:50 wt% 2N/DIO mixture in (**C,D**) a $d = 3.5$ μm thick, antipolar cell (with anti-parallel surface rubbing) and (**E**) a $d = 3.5$ μm thick, synpolar cell (with parallel surface rubbing), both in the absence of applied field. (**C**) The SmA_F phase grows in, upon slow cooling, from the top of this region of the cell at $T \approx 55$°C, irregular polygon-shaped domains of layers **n** and **P** oriented parallel to the cell plates and uniformly aligned throughout their volume. The existing N_F is in a surface-induced, π-twisted state, with **P** along the (antiparallel) buffing at the surfaces. This twisted state imposes no preferred bulk polarization orientation. As a result, the advancing SmA_F domains are ambivalent in their choice of polarization alignment and appear with **P** aligned locally either along +**z** or along -**z**, as shown. (**D**) A different part of the cell observed on heating from the SmA_F to the N_F phase. In their steady state, shown in *D1*, the SmA_F domains are generally extended along **z** to minimize polarization space-charge, with the domains separated by melted grain boundaries and polarization-

stabilized kinks, sketched in the inset in (*E*), which mediate small changes in the orientation of *P* along *z*. In contrast, non-zero ($\partial P_z/\partial y$) does not generate polarization charge, enabling neighboring domains with *P* directions that alternate in sign with changing *y*. Upon heating to the N$_F$, the boundaries between these adjacent domains transform into splay-bend walls (bright lines in *D2,3*), which then broaden into π-twist domains that eventually cover most of the cell (*D4,5*). (*E*) In the synpolar cell, a uniform monodomain is formed on cooling, with *n* in both the N$_F$ and SmA$_F$ phases generally along the buffing direction, giving excellent extinction, and *P* along the polar orientation preferred in the N$_F$ phase. The images show the texture around an air bubble extending through the thickness of the cell. The preferred orientation of *P* on the bubble boundary is tangential. At the bubble meridian, this boundary condition is compatible with the uniform polarization preferred by the cell surfaces but elsewhere, *n,P* twists in the interior of the cell to accommodate this boundary condition, and the cell has a non-extinguishing, yellow-green transmission color. This non-uniform state persists up to the dashed yellow lines, where the director field reverts to the preferred uniform state. These lines of polarization-stabilized kinks, globally parabolic in shape, having a local structure (shown in the inset) that minimizes polarization charge while mediating a change of orientation of *P*. Once the SmA$_F$ grows in, the expulsion of layer twist and bend forces more of the area surrounding the bubble into a uniform state, with the non-uniform region confined to a small area near the bubble. The residual pink transmission is presumably due to dislocations in the SmA$_F$ layering. Scale bars: (*C*) 500 μm; (*D*) 200 μm; (*E*) 100 μm.

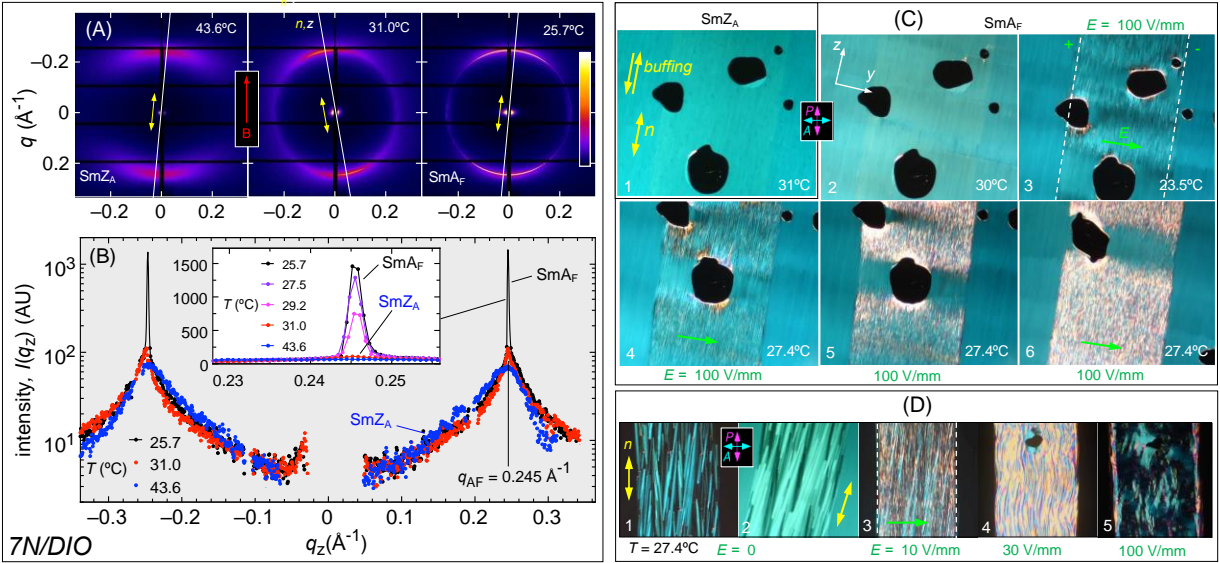

***Figure 3***: X-ray scattering and polarized microscopy textures of the $N_F$ and $SmA_F$ phases in the 50:50% 7N/DIO mixture. (***A***) Typical non-resonant SAXS obtained on cooling from $SmZ_A$ to $SmA_F$. In the $SmZ_A$ phase at 43.6°C, the SAXS shows a diffuse scattering arc, peaked along ***n*** at $q_z \sim 0.27$ Å$^{-1}$, from head-to-tail correlation of the mixture molecules, features also observed in the diffraction patterns of DIO [16]. (***B***) Radial intensity scans along the ***n,q_z*** direction (the white lines in (A)) at different temperatures. As in the 2N mixture, the scattering pattern rotates and spreads due to textural reorganization within the capillary as the $SmZ_A$ layers are replaced by $SmA_F$ layers. The scattering from the $SmZ_A$ layering is not visible here but is shown in ***Fig. 4***. Upon cooling, the diffuse peaks sharpen somewhat, the $SmZ_A$ layering peaks along $q_y$ weaken and disappear, and at $T \approx 31$°C, distinct, resolution-limited Bragg reflections appear along $q_z$ as shown in the inset, indicative of smectic ordering with the layer planes normal to ***n***. The scattering wavevector, $q_{zAF} \approx 0.245$ Å$^{-1}$, corresponds to a $SmA_F$ layer spacing of 25.6 Å, close to the wt% average molecular length of DIO (23.2 Å) and 7N (29.1 Å). The position of the $SmA_F$ scattering peak is very close to that of the diffuse nematic peak, as expected for an orthogonal smectic phase. (***C,D***) Polarized microscopy images of an antipolar cell with a $d = 3.5$ μm spacing and electrodes spaced by 1 mm (dashed white lines) for applying an in-plane field normal to the buffing direction, ***z***. The planar-aligned $SmZ_A$ texture shows only subtle changes upon transitioning to the $SmA_F$ (*C1,2*). This is because the antiparallel buffing, while it orients the director, does not favor either of the antiferroelectric polarization directions, so that at the transition the nanoscale antiferroelectric $SmZ_A$ layers normal to ***y*** simply coarsen into $SmA_F$ domains extended in ***z***, along the new layer normal, and alternating in polarization along ***y***. The director remains uniform through this change, giving a very similar appearance to the two phases. However, applying a small $E$-field applied along ***y*** (*C3-6*), causes the director in stripes of opposite ***P*** to rotate away from extinction in opposite directions, generating optical contrast that confirms their opposite polarity. The circular black regions are air bubbles, which effectively screen the applied electric field in the adjacent liquid

crystal, leaving the original texture undisturbed. (**D**) Annealing after such field treatment yields an inhomogeneous smectic fan texture (*D1,2*). In an applied field, these domains reorient, buckle, and, in sufficiently large applied field, coarsen to form large domains with the ***n***, ***z***, and ***P*** all oriented along the field, normal to the buffing direction (*D3 to 5*). This it appears that, during field-induced reorientation, ***n***, ***z***, and ***P*** remain coupled together, with the threshold originating from the elasticity and plasticity of the smectic layering. This threshold also results in the appearance of a coercive field in the polarization hysteresis (***Fig. 6***). The $N_F$ phase is readily reoriented by the weak stray applied fields over the electrodes. In the $SmA_F$ phase, however, there is a field threshold for such reorientation, so field effects are confined to the electrode gaps. Scale: the electrode gap (dashed white lines in *C*) is 1 mm wide.

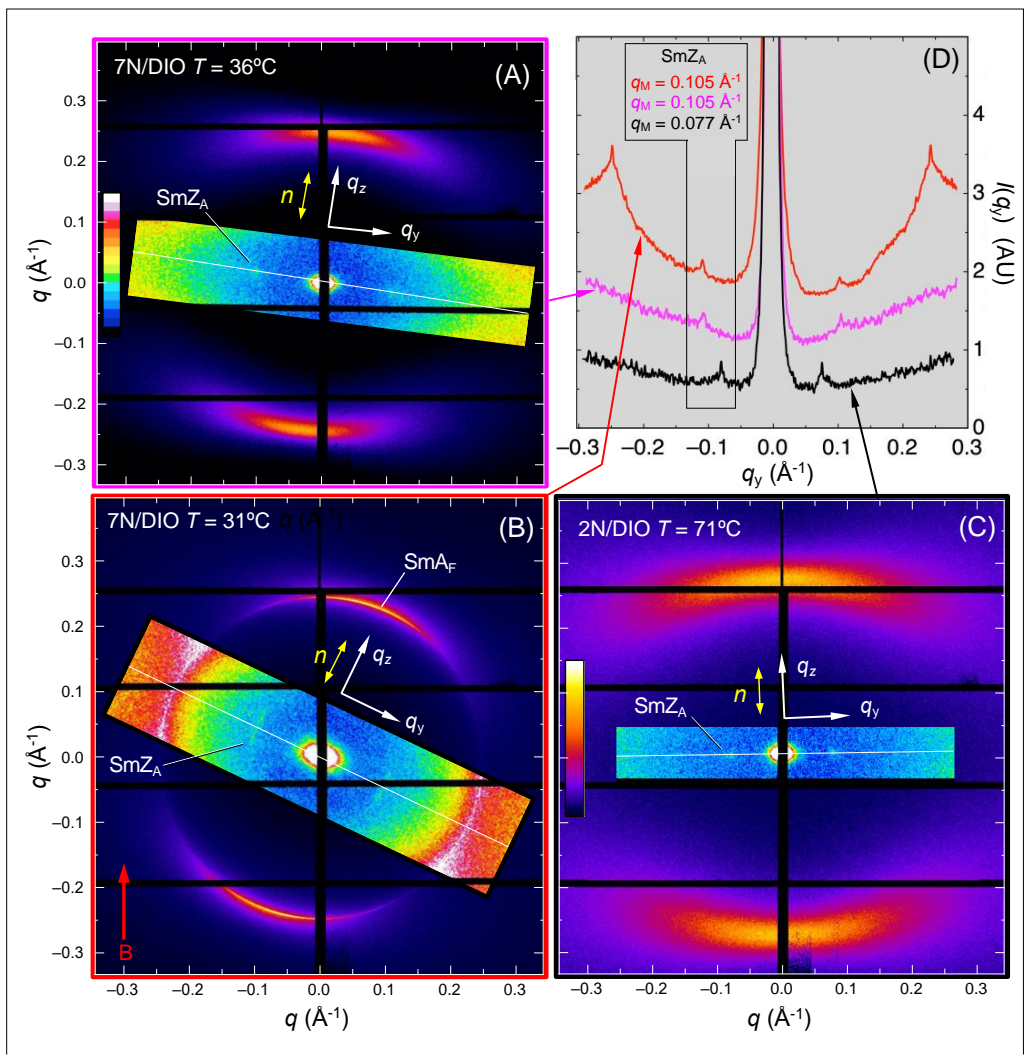

***Figure 4***: X-ray diffraction from the periodic density modulation of the SmZ$_A$ phase in the 7N/DIO and 2N/DIO mixtures. Panels (***A***) to (***C***) each show a complete SAXS image of the scattered intensity, $I(\boldsymbol{q})$, using the color gamut shown in (***C***). The rectangular overlays show $I(\boldsymbol{q})$ after histogram stretching and using the color gamut in (***B***), revealing the weak scattering peaks from the SmZ$_A$ layer modulation along $q_y$. The director is aligned in the nematic phase by a magnetic field, ***B***, but rearrangements of the sample in the capillary during cooling lead to some inhomogeneity of the SmZ$_A$ and SmA$_F$ layer orientation. (***A***) At $T$ = 36°C, the 7N/DIO mixture is in the SmZ$_A$ phase, as evidenced by the scattering along $\boldsymbol{q}_y$. The diffuse peaks along $\boldsymbol{q}_z$, parallel to the director, come from short-ranged, end-to-end molecular correlations. The SmZ$_A$ peak locations, at $|\boldsymbol{q}_y| = q_M \approx 0.105$ Å$^{-1}$, corresponds to a layer spacing of $d_M \approx 60$ Å, essentially independent of $T$. (***B***) Cooling to $T$ = 31°C initiates a weakly first-order phase transition to the SmA$_F$, with sharp scattering simultaneously from both the SmZ$_A$ and SmA$_F$ layers, indicating SmZ$_A$/SmA$_F$ phase coexistence. The SmZ$_A$ peaks at this temperature appear as extended arcs. The SmZ$_A$ scattering disappears ~ 0.5 °C below the onset of the SmZ$_A$ − SmA$_F$ transition, *i.e.*, there is a narrow range of $T$ where both the SmZ$_A$ and SmA$_F$ peaks are present, which we attribute to two phase coexistence at a first order transition. (**C**) Diffraction from the 2N/DIO mixture at $T = 71°C$, in the middle of the SmZ$_A$ phase region. (**D**) Radial scans of the scattered intensity along $q_y$, normal to the director, obtained by averaging $I(\boldsymbol{q})$ over the range of $q_z$ about $q_z = 0$ (white lines) that includes the SmZ$_A$ peaks ($\delta q_z \sim$ ±0.015 Å$^{-1}$). The low-temperature scan of (**B**) exhibits the SmZ$_A$ peaks at $q_y \approx 0.105$ Å$^{-1}$, as well as SmA$_F$ scattering at $q_y = 0.245$ Å$^{-1}$. While dwarfing the SmZ$_A$ peaks, this intensity is orders of magnitude smaller than the peak SmA$_F$ scattering along $q_z$.

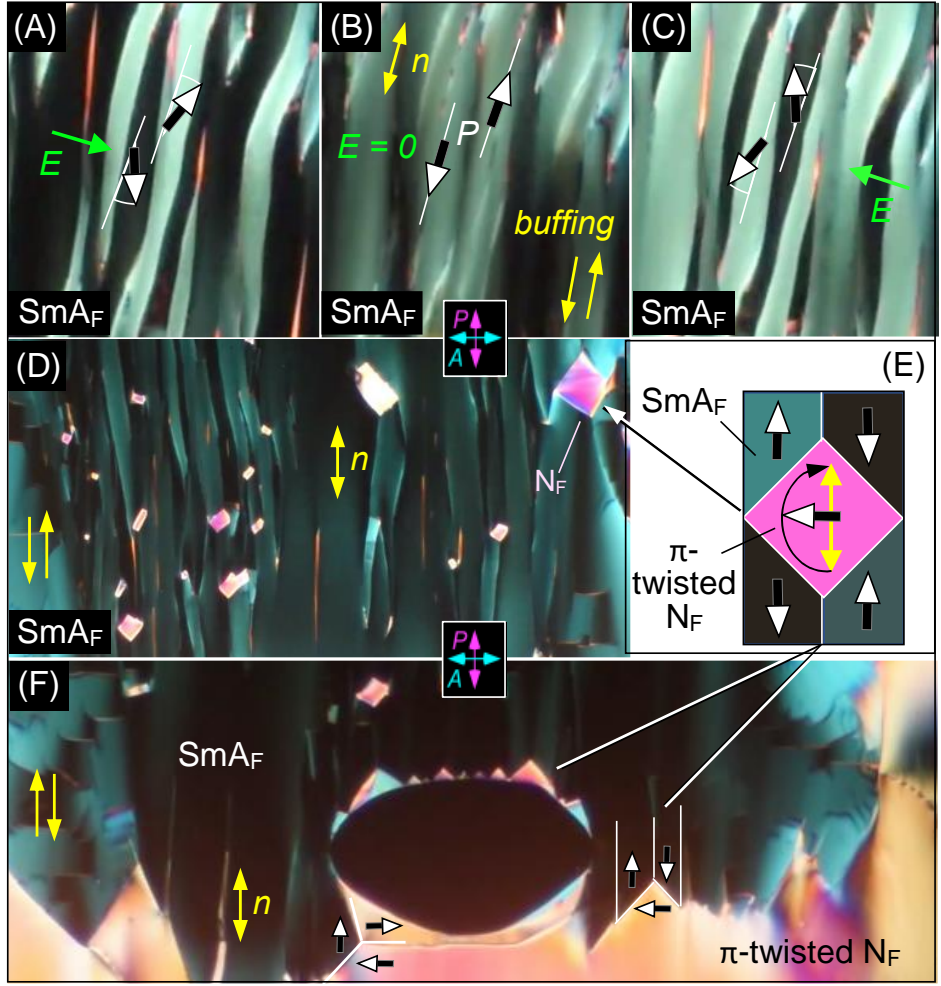

***Figure 5***: Response of SmA_F texture to field and frustration in the cell of ***Fig. 2 C,D*** ($d = 3.5$ μm thick cell with anti-parallel surface rubbing), in SmA_F domains (green) that have replaced a π-twisted N_F phase (yellow/pink). The twisted N_F structure does not bias the polarization preference so uniform domains of both signs of ***P*** should spontaneously appear in the SmA_F phase. (***A-C***) This can be tested by applying a transverse in-plane electric field ***E*** (substantially normal to ***n*** and ***P***) to a region of the cell with uniform, stripe-like domains. The applied field induces opposite rotations of the ***n-P*** couple in adjacent stripes, confirming that these are domains with opposing polarization. The sense of director rotation is reversed when the applied field direction is flipped. (***D,E***) Diamond-shaped inclusions of twisted N_F-phase material mediate the reversal (***D***) or termination of up-down pairs of SmA_F domains. The boundaries of these inclusions (***E***) are polarization-stabilized kinks, localized reorientations of ***P*** stabilized by the attraction of sheets of polarization charge of opposite sign.

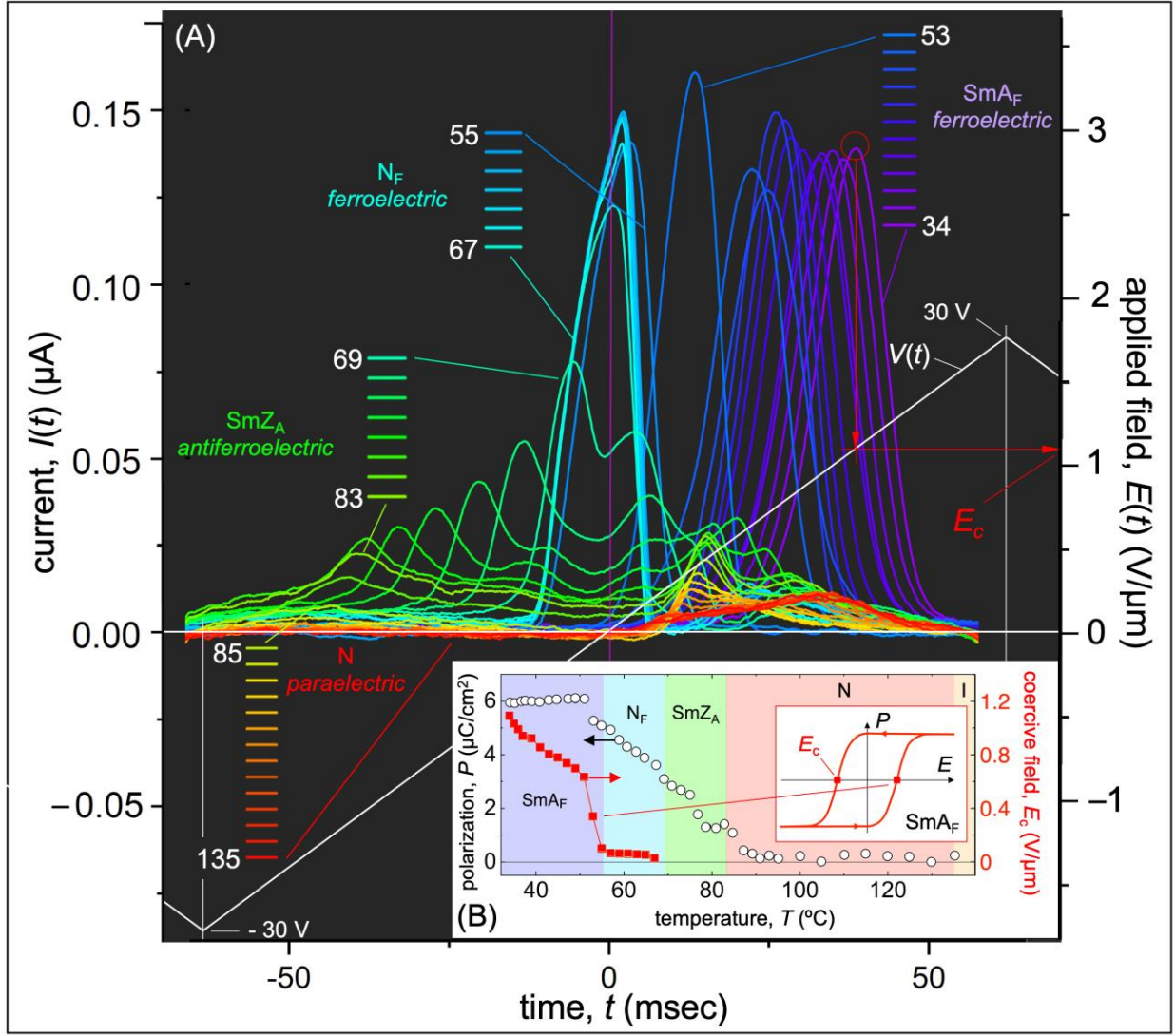

***Figure 6***: (**A**) *I(t)–V(t)* characteristics of the 2N/DIO mixture as a function of temperature with a 30 V peak amplitude, 8 Hz triangle-wave voltage (white trace) applied to a *d* = 17 μm ITO-sandwich cell with bookshelf layering in the smectic phases. The plot shows the current response during an N → SmZ$_A$ → N$_F$ → SmA$_F$ cooling scan. In the N phase (*T* > 84°C), the current shows only an ion peak following the sign change of *V(t)*. In the SmZ$_A$ phase (84°C > *T* > 68°C), two polarization peaks are seen during this half-cycle of the applied voltage, growing in area and occurring at smaller voltage on cooling. This is typical antiferroelectric behavior, the peaks marking the transition at finite voltage between the field-induced ferroelectric states and the equilibrium antiferroelectric state. In the N$_F$ phase, the Goldstone-mode mediated reorientation appears "thresholdless" and reversal of *P* produces a current peak at the zero-crossing of *V(t)*, followed by an ion peak for *t* > 0. The measured polarization in this phase is comparable to that of neat DIO. In the SmA$_F$ phase, the ion current disappears and polarization reversal peak occurs

at positive voltage corresponding to the coercive field $E_c$, shown, by way of example, for $T = 34°C$ (red construction). The temperature sequence of the $I(t)$ curves is 135, 130, 125, 120, 115, 110, 105, 100, 95, 93, 91, 89, 87, 85, 83, 81, 79, 77, 75, 73, 71, 69, 67, 65, 63, 61, 59, 57, 55, 53, 51, 49, 47, 45, 43, 41, 39, 38, 37, 35, and 34°C. (**B**) Polarization values $P(T)$ [open circles] were obtained by integrating the current. In the SmZ$_A$ phase, the polarization current generated following each zero-crossing of $V(t)$ overlaps with the ion current, so $P(T)$ is obtained in this case by doubling the area of the current peak generated before the zero-crossing. The coercive field, $E_c$, is also shown as a function of $T$ [solid symbols]. Note that in the N$_F$ and SmA$_F$ the ion peak is not observed because generally the electric field is very small in the in the ferroelectric phases because of screening by the polarization charge [28,29].